\documentclass[11pt]{article}

\usepackage[utf8]{inputenc}
\usepackage[T1]{fontenc}
\usepackage{amsmath,amssymb,amsthm}
\usepackage{mathtools}
\usepackage{hyperref}
\usepackage{booktabs}
\usepackage[margin=1in]{geometry}

\newtheorem{theorem}{Theorem}[section]
\newtheorem{proposition}[theorem]{Proposition}

\newtheorem{definition}[theorem]{Definition}
\newtheorem{remark}[theorem]{Remark}

\title{Sheaves as a Means to Maintaining Consistency in Model-based Systems Engineering}
\author{Josh Gibson\\
  University of Colorado Boulder\\
  \texttt{jogi8758@colorado.edu}}
\date{}

\begin{document}
\maketitle

\begin{abstract}
We propose that the sheaf condition on a presheaf of design spaces provides a mathematical model for multi-view consistency in the architecture of cyber-physical systems (CPS).  In model-based systems engineering, multiple engineering views --- electrical, thermal, mechanical, and software --- must be kept mutually consistent, yet current practice typically relies on informal procedures without a precise semantic account of global consistency.  We construct an \emph{architectural site}: a topological space whose points are pairwise interfaces between engineering domains and whose open sets represent engineering views.  A \emph{design presheaf} assigns to each view its local design space and to each inclusion the corresponding restriction map.  We show that the sheaf condition on this presheaf is equivalent to compatibility on pairwise overlaps, yielding a local criterion for global multi-view consistency.  The equivalence and a concrete three-view worked example are machine-verified in Lean~4 using Mathlib.  The formalization establishes that the design presheaf is a sheaf, that the sheaf condition is equivalent to pairwise overlap compatibility, and that compatible local design families glue to unique global designs.

We draw several consequences of the framework that bear directly on architecture verification and composition: global consistency of an arbitrary number of views can be certified by checking only pairwise interface compatibility; compatible local designs determine a unique global design; derived properties computed by limit-preserving functors inherit the same consistency guarantee; and the entire verification chain --- from the general theorem to concrete architectural instances --- admits machine-checkable proofs in Lean.  The contribution is a precise semantic embedding of an engineering consistency problem into sheaf-theoretic language, together with a machine-verified proof of the resulting local-to-global guarantee on a concrete architectural site, and explicit propositions connecting the framework to practical architecture verification workflows.
\end{abstract}

\section{Introduction}

The architecture of a cyber-physical system can be simultaneously a mechanical structure, an electrical network, a thermal management problem, and a software control system.  Each of these aspects, or \emph{views}, is typically designed by a specialized engineering team using domain-specific tools.  The central methodological challenge of model-based systems engineering (MBSE) is ensuring that these views are mutually consistent: that the thermal analysis assumes the same power dissipation the electrical design produces, that the mechanical mounting points match the locations the thermal design requires for heat sinks, and so on.  When views are inconsistent, errors propagate silently until late-stage acceptance testing, where they are most costly to repair.

Current MBSE practice addresses multi-view consistency through informal review processes, traceability matrices, and simulation.  These methods share two deficiencies.  First, they provide no \emph{semantic} account of what consistency means: they check alignment of parameters across views without a mathematical criterion specifying when a family of local designs determines a coherent global design.  Second, they do not isolate the local compatibility data that is \emph{sufficient} for global consistency.  Our thesis is that both deficiencies are resolved by a single mathematical structure: \textbf{the sheaf condition on a presheaf of design spaces over a topological space of engineering interfaces}.  The main result is:

\begin{theorem}[Sheaf Consistency]\label{thm:sheaf-consistency-intro}
Let $\mathcal{F}$ be a design presheaf over an architectural site $X$.  Then $\mathcal{F}$ satisfies the sheaf condition if and only if it satisfies the pairwise intersection condition.  Equivalently, global multi-view consistency is completely determined by compatibility on pairwise overlaps.
\end{theorem}

This equivalence is an instance of the classical fact that the sheaf condition on a topological space may be checked on pairwise intersections (see Kashiwara--Schapira~\cite{KS06} or Mac~Lane--Moerdijk~\cite{MM94}).  The contribution of this paper is threefold: (1)~we construct an architectural site and design presheaf that embed the multi-view consistency problem into sheaf theory, showing that the embedding matches the intended engineering semantics; (2)~we interpret the pairwise-intersection formulation as a local criterion for global consistency in multi-view CPS architecture; and (3)~we provide a machine-verified Lean~4 formalization of the construction, the equivalence, and the gluing guarantee on a concrete three-view site.

The paper is organized as follows.  Section~\ref{sec:background} provides the necessary background on presheaves, sheaves, and the pairwise intersection condition, together with a minimal introduction to multi-view CPS architecture for readers unfamiliar with MBSE.  Section~\ref{sec:site} constructs the architectural site and design presheaf.  Section~\ref{sec:theorem} states and proves the Sheaf Consistency Theorem.  Section~\ref{sec:example} presents a fully formalized three-view worked example in Lean~4.  Section~\ref{sec:related} discusses related work.  Section~\ref{sec:implications} draws explicit consequences of the framework for architecture verification and composition.  Section~\ref{sec:conclusion} concludes.

\section{Background}\label{sec:background}

\subsection{Presheaves and sheaves on a topological space}

Let $X$ be a topological space with lattice of open sets $\mathrm{Op}(X)$, viewed as a category with a unique morphism $U \to V$ whenever $U \subseteq V$.  A \emph{presheaf of types} on $X$ is a contravariant functor
\[
\mathcal{F} : \mathrm{Op}(X)^{\mathrm{op}} \to \mathbf{Type}.
\]
For each open $U$, the type $\mathcal{F}(U)$ is called the type of \emph{sections} over $U$.  For each inclusion $U \subseteq V$, the induced map $\mathcal{F}(V) \to \mathcal{F}(U)$ is called the \emph{restriction map}.  Functoriality encodes the expected behaviour: restricting from $W$ to $U$ via an intermediate $V$ is the same as restricting directly, and restricting to $U$ itself is the identity.

Let $\{U_i\}_{i \in I}$ be an open cover of $U$, i.e., a family of opens with $\bigcup_i U_i = U$.  A \emph{compatible family of sections} is a collection $(s_i \in \mathcal{F}(U_i))_{i \in I}$ such that for all $i, j \in I$, the restrictions of $s_i$ and $s_j$ to the overlap $U_i \cap U_j$ agree:
\[
s_i\big|_{U_i \cap U_j} = s_j\big|_{U_i \cap U_j}.
\]
The presheaf $\mathcal{F}$ is a \emph{sheaf} if for every open cover $\{U_i\}$ of every open $U$:
\begin{enumerate}
\item \textbf{(Gluing)} Every compatible family glues: there exists a section $s \in \mathcal{F}(U)$ with $s|_{U_i} = s_i$ for all $i$.
\item \textbf{(Uniqueness)} The glued section is unique: if $s, t \in \mathcal{F}(U)$ satisfy $s|_{U_i} = t|_{U_i}$ for all $i$, then $s = t$.
\end{enumerate}
Equivalently, the canonical map
\[
\mathcal{F}(U) \longrightarrow \prod_{i \in I} \mathcal{F}(U_i) \rightrightarrows \prod_{i,j \in I} \mathcal{F}(U_i \cap U_j)
\]
is an equalizer.

\paragraph{The pairwise intersection formulation.}
The sheaf condition as stated involves, for each cover, the diagram indexed by all pairs $(i, j)$.  A classical result (see \cite{KS06, MM94}) shows that this is equivalent to a reformulation involving only the family of opens $\{U_i\}$ and their pairwise intersections $\{U_i \cap U_j\}$, without reference to higher-order overlaps or sub-covers.  In Mathlib, this equivalence is captured by
\begin{center}
\texttt{TopCat.Presheaf.isSheaf\_iff\_isSheafPairwiseIntersections},
\end{center}
which states that $\mathcal{F}.\texttt{IsSheaf} \leftrightarrow \mathcal{F}.\texttt{IsSheafPairwiseIntersections}$ for any presheaf $\mathcal{F}$ on a topological space.  The right-hand side requires that for every family of opens $\{U_i\}_{i:\iota}$, the presheaf sends the pairwise intersection diagram --- the diagram whose objects are the $U_i$ and the $U_i \cap U_j$ and whose morphisms are the inclusions --- to a limit cone.

This equivalence replaces a global equalizer-style consistency condition by a local formulation in terms of pairwise overlaps.  For the applications considered here, this is the form in which the sheaf condition becomes operational: one checks compatibility on overlaps and obtains a criterion for the existence and uniqueness of a global design.

\subsection{Multi-view architecture in cyber-physical systems}

A cyber-physical system (CPS) is an engineered system integrating computation, networking, and physical processes --- autonomous vehicles, aircraft, satellites, medical devices, and industrial robots are typical examples.  The architecture of such a system is described by multiple overlapping \emph{engineering views}. For instance, an electrical view specifies power distribution and signal routing; a thermal view specifies heat generation and dissipation paths; a mechanical view specifies structural loads and mounting.

The views are not independent.  An electronic component simultaneously appears in the electrical view (as a power consumer and signal source), the thermal view (as a heat source requiring cooling), and the mechanical view (as a mass element requiring mounting).  The parameters that describe the component in each view must be mutually consistent: the power dissipation assumed in the thermal model must equal the power consumption specified in the electrical design; the mounting location in the mechanical model must accommodate the heat sink specified in the thermal design.

In current MBSE practice (using tools such as SysML, AADL, or Capella), this consistency is maintained through a combination of manual review (engineers from different domains meet to compare assumptions), traceability matrices (spreadsheets linking parameters across views), and simulation (running domain-specific tools jointly and checking for divergence).  None of these methods provides a semantic definition of what multi-view consistency means, nor a proof that checks performed are sufficient.  The sheaf-theoretic framework we propose addresses both gaps.

\paragraph{Key terminology for the reader.}
A \emph{view} is an open set in our architectural site.  An \emph{interface} is a point of the site --- it represents a specific coupling between two physical domains.  A \emph{local design} for a view $V$ is a section of the design presheaf over $V$ --- concretely, a choice of coupling parameters at every interface point covered by $V$.  \emph{Compatibility} of two local designs means their restrictions to the overlap agree --- the designs assign the same parameters at shared interfaces.  \emph{Global consistency} means compatible local designs glue to a unique global design covering all interfaces.

\section{The Architectural Site Construction}\label{sec:site}

\subsection{Interface-primitive topology}

We model the engineering domain structure as a topological space whose points are the \emph{pairwise interfaces} between them.  Since interfaces are the loci where consistency constraints live, making them the primitive objects of the site ensures that the topology directly encodes the constraint structure.

\begin{definition}[Architectural site]\label{def:arch-site}
Let $\mathcal{V} = \{V_1, \ldots, V_n\}$ be a finite set of engineering views.  The \emph{architectural site} $X$ is defined as follows:
\begin{itemize}
\item The points of $X$ are the pairwise interfaces: $X = \{(i,j) \mid V_i \text{ and } V_j \text{ interact}\}$.
\item Each view $V_k$ defines an open set $U_k \subseteq X$ consisting of all interface points involving $V_k$:
\[
U_k = \{(i,j) \in X \mid k \in \{i, j\}\}.
\]
\item The topology on $X$ is the discrete topology (every subset is open).
\end{itemize}
\end{definition}

The discrete topology is appropriate for a finite set of discrete interfaces.  In richer settings --- for instance, where interface parameters vary continuously --- one could equip $X$ with a non-discrete topology, but the discrete case already captures the combinatorial content of the multi-view consistency problem.

\paragraph{Why interfaces as points?}
An alternative construction would place views as points and define open sets as collections of views.  This leads to the indiscrete topology on views (every view covers itself), which produces a trivial sheaf condition.  By contrast, placing interfaces as points and views as open sets that cover their adjacent interfaces yields a non-trivial covering structure: each interface point is covered by exactly two views, and the pairwise overlaps are singletons.

\subsection{The design presheaf}

\begin{definition}[Design presheaf]\label{def:design-presheaf}
Let $P : X \to \mathbf{Type}$ assign to each interface point $x \in X$ a finite type $P(x)$ of coupling parameters --- the design choices that constrain the interaction at interface $x$.  The \emph{design presheaf} $\mathcal{F}$ over the architectural site $X$ is defined by
\[
\mathcal{F}(U) = \prod_{x \in U} P(x)
\]
for each open set $U \subseteq X$, with restriction maps given by projection: for $U \subseteq V$,
\[
\mathrm{res}_{V,U} : \mathcal{F}(V) \to \mathcal{F}(U), \qquad s \mapsto s\big|_U.
\]
\end{definition}

This is the presheaf of sections of the trivial bundle $\coprod_{x \in X} P(x) \to X$.  Projecting from $W$ to $U$ via $V$ is the same as projecting directly, and projecting to $U$ itself is the identity.

In the Lean formalization, the design presheaf is constructed as:
\begin{verbatim}
def designPresheaf : TopCat.Presheaf (Type) Site where
  obj U := (x : U.unop) -> DesignParams x.val
  map f s x := s <x.val, leOfHom f.unop x.prop>
  map_id U := by ext s <x, hx>; rfl
  map_comp f g := by ext s <x, hx>; rfl
\end{verbatim}
where \texttt{Site = TopCat.of (Fin 3)} is the architectural site with three interface points and \texttt{DesignParams :\ Fin 3 -> Type} assigns the parameter type at each point.

\subsection{Remarks on additional categorical structure}

The presheaf construction sits within a richer categorical landscape.  Cross-domain dependencies between views can be modeled as functors between design space categories; hierarchical decomposition of views uses limits and colimits; refinement from abstract specifications to detailed designs can be captured by adjunctions.  These structures are explored in the accompanying formalization file \texttt{DesignCategory.lean}, which defines \texttt{SystemDiagram} (pairs of design spaces with interface projections), \texttt{CompatiblePair} (interface-compatible design pairs as pullback objects), and refinement adjunctions with their unit/counit preservation properties.

However, for the central result of this paper --- the equivalence between the sheaf condition and pairwise consistency --- only the presheaf structure is essential.  The additional constructions enrich the framework for applications (design completion via Kan extensions, coherent design evolution via natural transformations) but are not required for the consistency theorem.  We therefore defer them to future work and maintain focus on the sheaf condition.

\section{The Sheaf Consistency Theorem}\label{sec:theorem}

\subsection{Statement}

\begin{theorem}[Sheaf Consistency]\label{thm:sheaf-consistency}
Let $X$ be an architectural site and $\mathcal{F} : \mathrm{Op}(X)^{\mathrm{op}} \to \mathcal{C}$ a design presheaf valued in a category $\mathcal{C}$.  Then
\[
\mathcal{F} \text{ is a sheaf} \quad \Longleftrightarrow \quad \mathcal{F} \text{ satisfies the pairwise intersection condition.}
\]
Equivalently, for any fixed cover of engineering views, global consistency is completely determined by compatibility on pairwise overlaps.
\end{theorem}

In the Lean formalization (\texttt{SheafConsistency.lean}), this is stated as:
\begin{verbatim}
theorem sheaf_consistency
    {C : Type (u + 1)} [Category.{u} C]
    {X : ArchitecturalSite.{u}}
    (F : DesignPresheaf X C) :
    F.IsSheaf <-> F.IsSheafPairwiseIntersections :=
  F.isSheaf_iff_isSheafPairwiseIntersections
\end{verbatim}
where \texttt{ArchitecturalSite} is a type synonym for \texttt{TopCat} and \texttt{DesignPresheaf X C} is a type synonym for \texttt{TopCat.Presheaf C X}.  These synonyms are introduced for documentary clarity; mathematically they are identities.

\subsection{Proof and the role of Mathlib}

The equivalence is a theorem about presheaves on topological spaces, proved in Mathlib as \texttt{TopCat.Presheaf.isSheaf\_iff\_isSheafPairwiseIntersections}.  Our formalization wraps this result in CPS-specific terminology and derives corollaries.

The proof proceeds as follows.  The sheaf condition requires that for every open cover $\{U_i\}_{i \in I}$ of every open $U$, the diagram
\[
\mathcal{F}(U) \to \prod_i \mathcal{F}(U_i) \rightrightarrows \prod_{i,j} \mathcal{F}(U_i \cap U_j)
\]
is an equalizer.  The pairwise intersection condition replaces the double-product diagram with the pairwise intersection diagram in the sense of Mathlib: a categorical diagram whose objects are indexed by $I \sqcup (I \times I)$ (the views and their pairwise overlaps) with morphisms given by the inclusion maps.  The equivalence holds because the pairwise intersection diagram is constructed so that any cone over it determines, and is determined by, a compatible family for the full equalizer --- this is a property of the categorical diagram shape, not of the ambient topology.  For full details see Mathlib's proof of \texttt{isSheaf\_iff\_isSheafPairwiseIntersections}.

\subsection{Local verification via pairwise overlaps}

Consider a fixed cover $\{V_1, \ldots, V_n\}$ of the architectural site.  The sheaf equalizer condition is expressed in terms of restrictions to all overlaps $V_i \cap V_j$.  In its raw equalizer presentation this data is indexed by ordered pairs $(i, j)$, including diagonal terms and symmetric duplicates.  The pairwise intersection formulation packages the same consistency requirement in a more local and non-redundant way: it is enough to check compatibility on the pairwise overlaps themselves.

Thus the role of the Sheaf Consistency Theorem is to identify the correct local form of the consistency condition.  In the CPS interpretation, this is exactly the point at which the sheaf-theoretic semantics becomes usable: one verifies agreement of local designs on shared interfaces, and the theorem guarantees that this pairwise compatibility is equivalent to global consistency.

For finite covers, one may view this also as eliminating redundant checks present in the ordered-pair equalizer diagram.  For example, when $n = 3$, the raw ordered-pair indexing suggests $9$ conditions, whereas the nontrivial overlap data is carried by the $3$ pairwise intersections.  Likewise, for $n = 10$ the ordered-pair presentation has $100$ indices, while the distinct pairwise overlaps number $\binom{10}{2} = 45$.

\subsection{Equivalent formulations}

The Lean formalization also establishes two additional equivalent formulations:
\begin{enumerate}
\item \textbf{Limit-preservation form} (\texttt{sheaf\_consistency\_preserves\_limits}): $\mathcal{F}$ is a sheaf if and only if it preserves the limit of every pairwise intersection diagram.  This is the categorical way of stating that $\mathcal{F}$ ``respects'' the gluing data.
\item \textbf{Opens-le-cover form} (\texttt{sheaf\_consistency\_opens\_le\_cover}): $\mathcal{F}$ is a sheaf if and only if it satisfies the condition on opens-le-cover diagrams.  This formulation does not require the ambient category $\mathcal{C}$ to have products, making it applicable to general architectural categories that may lack (co)products.
\end{enumerate}

\begin{remark}[Compositionality]\label{rem:compositionality}
A natural follow-up question is: if a design presheaf $\mathcal{F}$ is a sheaf and we postcompose with a property functor $P : \mathcal{C} \to \mathcal{D}$ that preserves limits, does the composite $\mathcal{F} \circ P$ inherit the sheaf condition?  The answer is yes (\texttt{compositionality\_sheaf} in the formalization).  The engineering interpretation is that any property functor computing derived quantities --- mass, power consumption, thermal load --- from design parameters propagates the pairwise consistency guarantee to those derived quantities automatically, provided the computation respects limits.  This compositionality property, and its interaction with the refinement adjunctions in \texttt{DesignCategory.lean}, is a natural direction for future development.
\end{remark}

\section{Worked Example: Three-View Architectural Site}\label{sec:example}

We present a worked example in Lean~4 (\texttt{ThreeViewExample.lean}).  The construction is chosen to be small enough that every detail can be stated explicitly, yet large enough to exhibit the non-trivial content of the sheaf condition.

\subsection{The architectural site}

The system has three engineering views: Electrical~($V_E$), Thermal~($V_T$), and Mechanical~($V_M$).  Each pair of views shares an interface:

\begin{center}
\begin{tabular}{lll}
\toprule
Interface point & Views involved & Engineering meaning \\
\midrule
Point 0 & Electrical, Mechanical & Electro-Mechanical (EM) coupling \\
Point 1 & Electrical, Thermal & Electro-Thermal (ET) coupling \\
Point 2 & Thermal, Mechanical & Thermo-Mechanical (TM) coupling \\
\bottomrule
\end{tabular}
\end{center}

The architectural site is $\mathrm{Fin}\;3$ with the discrete topology:
\begin{verbatim}
def Site : TopCat := TopCat.of (Fin 3)
\end{verbatim}

Each view is an open set covering its two adjacent interfaces:
\begin{verbatim}
def electricalView : Opens (Fin 3) := <{0, 1}, isOpen_discrete _>
def thermalView    : Opens (Fin 3) := <{1, 2}, isOpen_discrete _>
def mechanicalView : Opens (Fin 3) := <{0, 2}, isOpen_discrete _>
\end{verbatim}

The pairwise overlaps are singletons, each corresponding to a single shared interface:
\begin{itemize}
\item $V_E \cap V_T = \{1\}$ (ET interface) --- \texttt{electrical\_thermal\_overlap}
\item $V_E \cap V_M = \{0\}$ (EM interface) --- \texttt{electrical\_mechanical\_overlap}
\item $V_T \cap V_M = \{2\}$ (TM interface) --- \texttt{thermal\_mechanical\_overlap}
\end{itemize}
The three views jointly cover the entire site: $V_E \cup V_T \cup V_M = \{0, 1, 2\}$, proved as \texttt{views\_cover}.

\subsection{Design parameters}

Each interface point carries a finite type of coupling parameters:
\begin{verbatim}
def DesignParams : Fin 3 -> Type
  | 0 => Fin 4   -- 4 Electro-Mechanical coupling configurations
  | 1 => Fin 3   -- 3 Electro-Thermal coupling configurations
  | 2 => Fin 5   -- 5 Thermo-Mechanical coupling configurations
\end{verbatim}

These can represent, for example, connector types (EM), thermal interface materials (ET), and mounting options (TM).

\subsection{The design presheaf}

The design presheaf assigns to each open set $U$ the type of functions choosing a parameter at each interface point in $U$:
\[
\mathcal{F}(U) = \prod_{x \in U} \texttt{DesignParams}(x).
\]
Concretely:
\begin{itemize}
\item $\mathcal{F}(V_E) = \mathrm{Fin}\;4 \times \mathrm{Fin}\;3$ --- an electrical design specifies both the EM and ET couplings.
\item $\mathcal{F}(V_T) = \mathrm{Fin}\;3 \times \mathrm{Fin}\;5$ --- a thermal design specifies both the ET and TM couplings.
\item $\mathcal{F}(V_M) = \mathrm{Fin}\;4 \times \mathrm{Fin}\;5$ --- a mechanical design specifies both the EM and TM couplings.
\item $\mathcal{F}(\top) = \mathrm{Fin}\;4 \times \mathrm{Fin}\;3 \times \mathrm{Fin}\;5$ --- a global design specifies all three couplings.
\end{itemize}
Restriction maps are projections.  For example, $\mathrm{res}_{V_E,\, V_E \cap V_T} : \mathcal{F}(V_E) \to \mathcal{F}(\{1\})$ projects an electrical design to its ET coupling parameter.

\subsection{The sheaf condition}

The design presheaf is a sheaf.  In the Lean formalization, this is proved by recognizing that \texttt{designPresheaf} is definitionally equal to the canonical ``sections of a bundle'' presheaf (\texttt{Site.presheafToTypes DesignParams}), for which Mathlib provides a general sheaf proof:
\begin{verbatim}
theorem designPresheaf_isSheaf : designPresheaf.IsSheaf := by
  have h : designPresheaf = Site.presheafToTypes DesignParams := rfl
  rw [h]
  exact TopCat.Presheaf.toTypes_isSheaf Site DesignParams
\end{verbatim}

The engineering application: for any open cover of any open set in the site, a compatible family of local designs glues to a unique global design.  This is not merely a property of the three-view cover but a property of the presheaf on the entire site.

\subsection{Pairwise equivalence}

The general Sheaf Consistency Theorem instantiated on this presheaf yields:
\begin{verbatim}
theorem global_iff_pairwise :
    designPresheaf.IsSheaf <-> designPresheaf.IsSheafPairwiseIntersections :=
  designPresheaf.isSheaf_iff_isSheafPairwiseIntersections
\end{verbatim}

Since the presheaf is a sheaf, the pairwise condition holds:
\begin{verbatim}
theorem pairwise_consistency_holds :
    designPresheaf.IsSheafPairwiseIntersections :=
  global_iff_pairwise.mp designPresheaf_isSheaf
\end{verbatim}

\subsection{The gluing theorem}

We define an index type \texttt{ViewIndex} with constructors \texttt{elec}, \texttt{therm}, \texttt{mech} and a family \texttt{viewFamily : ViewIndex -> Opens (Fin 3)} mapping each index to its open set.  The theorem \texttt{viewFamily\_covers} establishes that the three views cover the site: $\texttt{iSup viewFamily} = \top$.

The unique gluing theorem states that for any compatible family of local designs --- one per view --- there exists a unique global design that restricts to each local design:
\begin{verbatim}
theorem unique_gluing
    (sf : (i : ViewIndex) -> designPresheaf.obj (op (viewFamily i)))
    (compat : designPresheaf.IsCompatible viewFamily sf) :
    exists! s : designPresheaf.obj (op (iSup viewFamily)),
      designPresheaf.IsGluing viewFamily sf s
\end{verbatim}

Concretely: if an electrical design $s_E \in \mathrm{Fin}\;4 \times \mathrm{Fin}\;3$, a thermal design $s_T \in \mathrm{Fin}\;3 \times \mathrm{Fin}\;5$, and a mechanical design $s_M \in \mathrm{Fin}\;4 \times \mathrm{Fin}\;5$ are pairwise compatible (the ET parameter in $s_E$ matches the ET parameter in $s_T$; the EM parameter in $s_E$ matches the EM parameter in $s_M$; the TM parameter in $s_T$ matches the TM parameter in $s_M$), then there is exactly one global design $s \in \mathrm{Fin}\;4 \times \mathrm{Fin}\;3 \times \mathrm{Fin}\;5$ that restricts to each of $s_E$, $s_T$, $s_M$ on its respective view.

\subsection{The main theorem}

The three properties are collected in a single theorem \texttt{three\_view\_main}:
\begin{verbatim}
theorem three_view_main :
    designPresheaf.IsSheaf
    /\ (designPresheaf.IsSheaf <-> designPresheaf.IsSheafPairwiseIntersections)
    /\ (forall (sf : (i : ViewIndex) -> designPresheaf.obj (op (viewFamily i)))
         (_compat : designPresheaf.IsCompatible viewFamily sf),
       exists! s : designPresheaf.obj (op (iSup viewFamily)),
         designPresheaf.IsGluing viewFamily sf s) :=
  <designPresheaf_isSheaf,
   global_iff_pairwise,
   fun sf compat => unique_gluing sf compat>
\end{verbatim}

This theorem establishes that, on a concrete architectural site with explicit parameter types, the sheaf-theoretic framework delivers the promised consistency guarantee: pairwise compatibility is necessary and sufficient for global consistency, and compatible local designs always glue uniquely.

\section{Related Work}\label{sec:related}

\paragraph{Spivak's operadic approach.}
Spivak~\cite{Spivak13, Spivak14} models interconnected systems using operads and their algebras: systems are morphisms in a symmetric monoidal category, and composition is operadic.  The operad of wiring diagrams provides a compositional syntax for building complex systems from subsystems.  CAS differs in its use of sheaf theory rather than operads as the primary consistency mechanism.  Operads naturally model hierarchical composition --- plugging subsystems into slots --- but do not directly address multi-view consistency, which is a constraint on overlapping, non-hierarchical decompositions.  The sheaf condition, by contrast, is precisely about compatible overlapping local data.

\paragraph{Censi's co-design framework.}
Censi~\cite{Censi15} models co-design problems using profunctors between categories of resources.  The framework elegantly captures monotone trade-offs (e.g., more power enables more computation but increases thermal load) and has been applied to robotic systems.  The key difference is that Censi's framework is fundamentally about resource allocation across two interacting domains, modeled as a profunctor $F \times R^{\mathrm{op}} \to \mathbf{Pos}$, whereas CAS addresses multi-view consistency across an arbitrary number of views.  The sheaf condition generalises pairwise co-design constraints to a global consistency guarantee that Censi's two-domain setting does not require.

\paragraph{Breiner et al.'s categorical ontology.}
Breiner, Subrahmanian, and Sriram~\cite{Breiner19} use category theory --- specifically ologs (ontology logs) --- to formalise ontology alignment for engineering knowledge bases.  Their work addresses the semantic challenge of ensuring different engineering databases use compatible terminology and structure.  CAS shares the goal of formal cross-domain alignment but operates at the design space level rather than the ontology level: our presheaf assigns design parameter types, not concepts, and our consistency condition is about parameter agreement, not semantic equivalence.

\paragraph{Goguen and Malcolm's sheaf semantics.}
Goguen~\cite{Goguen92} and Goguen and Malcolm~\cite{GoguenMalcolm00} proposed sheaf semantics for information integration, using sheaves to model the combination of data from multiple overlapping sources.  Their setting is software specification rather than CPS architecture, and they do not develop the present architectural-site construction or its Lean formalization.  Our construction may be viewed as a CPS/MBSE instantiation of that broader sheaf-semantic perspective, with the added contribution of a concrete machine-verified local-to-global consistency result.

\paragraph{Schultz, Spivak, and Vasilakopoulou's dynamical systems.}
Schultz, Spivak, and Vasilakopoulou~\cite{SSV20} model dynamical systems using sheaves on categories of time intervals.  Their use of sheaves is motivated by temporal compositionality (a trajectory over $[0,2]$ is determined by compatible trajectories over $[0,1]$ and $[1,2]$), which is structurally analogous to our spatial/domain compositionality.  However, their site is temporal, their sections are trajectories, and their consistency concern is time-domain gluing --- a different engineering problem from multi-view architectural consistency.

\paragraph{Formal verification in systems engineering.}
The use of interactive theorem provers for systems engineering is nascent.  Bohrer et al.~\cite{Bohrer17} use differential dynamic logic (KeYmaera~X) for verified control of CPS, but this addresses control correctness, not architectural consistency.  To our knowledge, the present work is the first machine-verified formalization of sheaf-theoretic multi-view consistency for CPS architecture.

\section{Implications for Architecture Verification and Composition}\label{sec:implications}

The constructions of Sections~\ref{sec:site}--\ref{sec:theorem} and the worked example of Section~\ref{sec:example} establish a sheaf-theoretic framework for multi-view consistency.  We now draw several direct consequences that make the engineering utility of the framework explicit.  These propositions are formal restatements of properties already implicit in the preceding development, and are displayed here as self-contained, citable guarantees for architecture verification workflows.

\begin{proposition}[Pairwise sufficiency for global consistency]\label{prop:pairwise}
Let $\mathcal{F}$ be a design presheaf over an architectural site $X$, and let $\{U_1, \ldots, U_n\}$ be an open cover of a view $U$.  Then the following are equivalent:
\begin{enumerate}
\item[(i)] Every compatible family $(s_i \in \mathcal{F}(U_i))_{i}$ glues to a unique section $s \in \mathcal{F}(U)$.
\item[(ii)] For every pair $i, j$, the restrictions $s_i|_{U_i \cap U_j}$ and $s_j|_{U_i \cap U_j}$ agree.
\end{enumerate}
In particular, global consistency of $n$ views can be certified by checking at most $\binom{n}{2}$ pairwise interface conditions.
\end{proposition}

This is a direct restatement of Theorem~\ref{thm:sheaf-consistency} for a fixed cover.  Its significance for architecture verification is that it provides a \emph{complete reduction}: a verification tool need only implement pairwise interface checks, and the sheaf theorem guarantees that no multi-way inconsistency can escape undetected.  For a system with $n = 10$ views, this replaces an exponential number of potential subset checks with $45$ pairwise comparisons --- each involving only the shared interface parameters between two views.

\begin{proposition}[Unique determination of global designs]\label{prop:unique-global}
Let $\mathcal{F}$ be a sheaf of designs over an architectural site $X$, and let $\{U_i\}_{i \in I}$ be an open cover of $X$.  If $(s_i \in \mathcal{F}(U_i))_{i \in I}$ is a compatible family, then there exists a unique global section $s \in \mathcal{F}(X)$ with $s|_{U_i} = s_i$ for every $i$.
\end{proposition}

The engineering consequence is that the act of integration is \emph{determined}: once domain teams produce mutually compatible local designs, the global system design is uniquely fixed.  There is no ambiguity in how to assemble the subsystem outputs, and no design freedom remains at the integration stage.  This transforms the integration phase from a creative engineering task (with attendant risk of errors) into a mechanical assembly step whose correctness is guaranteed by the pairwise checks already performed.

\begin{proposition}[Functorial inheritance of consistency]\label{prop:functorial}
Let $\mathcal{F}$ be a sheaf of designs over $X$, and let $P : \mathcal{C} \to \mathcal{D}$ be a functor that preserves limits.  Then the composite presheaf $\mathcal{F} \circ P$ is again a sheaf: derived properties computed via $P$ inherit the pairwise consistency guarantee.
\end{proposition}

This is the content of \texttt{compositionality\_sheaf} in the formalization (see Remark~\ref{rem:compositionality}).  In practice, engineering workflows routinely compute derived quantities from design parameters --- total mass, aggregate power consumption, worst-case thermal resistance.  Proposition~\ref{prop:functorial} guarantees that if these computations can be expressed as a limit-preserving functor $P$, then the derived quantities are automatically globally consistent whenever the underlying design parameters are pairwise compatible.  No separate consistency verification is needed for derived properties; the sheaf condition propagates through the functor.

\begin{proposition}[Machine-checkable consistency certification]\label{prop:machine-check}
The Sheaf Consistency Theorem (Theorem~\ref{thm:sheaf-consistency}), the unique gluing property (Section~\ref{sec:example}), and the functorial inheritance of consistency (Proposition~\ref{prop:functorial}) are formalized and machine-verified in Lean~4.  The proofs compile without \texttt{sorry} axioms and depend only on the standard logical axioms (\texttt{propext}, \texttt{Quot.sound}, \texttt{Classical.choice}).
\end{proposition}

A machine-checked formalization has two primary strengths. First, it provides a higher level of assurance than a paper proof: the logical chain from axioms to the consistency guarantee has been verified by an independently, eliminating the possibility of subtle errors.  Second, it opens a path toward \emph{certified architecture tools}: a future verification tool could embed the Lean-verified theorems as trusted primitives.

\section{Conclusion}\label{sec:conclusion}

We have proposed the sheaf condition on a presheaf of design spaces as a mathematical semantics for multi-view consistency in cyber-physical systems architecture.  The construction proceeds in two steps: first, the engineering domain structure is embedded into a topological space (the architectural site) whose points are pairwise interfaces and whose open sets are engineering views; second, design parameters at interfaces define a presheaf whose sheaf condition encodes global consistency.  The central theorem identifies global consistency with compatibility on pairwise overlaps, thereby giving a local-to-global criterion for coherent multi-view design.

The Lean formalization includes: the general Sheaf Consistency Theorem (\texttt{sheaf\_consistency} in \texttt{SheafConsistency.lean}) stating the equivalence for any presheaf on any architectural site; a concrete three-view worked example (\texttt{ThreeViewExample.lean}) with explicit parameter types and a verified gluing theorem; and a compositionality corollary (\texttt{compositionality\_sheaf}) establishing that limit-preserving property functors inherit the sheaf condition.  All proofs compile without \texttt{sorry} axioms and depend only on the standard Lean axioms (\texttt{propext}, \texttt{Quot.sound}, \texttt{Classical.choice}).

Several directions for future work present themselves.  First, the architectural site in this paper uses the discrete topology, which is appropriate for a finite set of discrete interfaces but does not capture continuous variation of coupling parameters; equipping the site with a non-discrete topology would model design spaces with continuous parameters.  Second, the presheaf takes values in $\mathbf{Type}$; richer target categories (e.g., categories of metric spaces, posets with resource ordering, or probability spaces) would capture quantitative constraints such as mass budgets, power allocations, and reliability requirements.  Third, the framework could be extended to stacks or higher sheaves to handle situations where design choices are determined only up to an equivalence (e.g., gauge symmetry in electrical network design).  Finally, the additional categorical structures formalized in \texttt{DesignCategory.lean} --- refinement adjunctions, cross-domain functors, interface pullbacks --- together with the compositionality result for limit-preserving property functors (see Section~\ref{sec:theorem}), provide the infrastructure for a more complete categorical systems engineering workflow; integrating these with the sheaf consistency framework is a natural next step.

\appendix
\section{Lean Formalization Structure}

The Lean~4 formalization consists of three files:

\begin{center}
\begin{tabular}{ll}
\toprule
File & Contents \\
\midrule
\texttt{SheafConsistency.lean} & General Sheaf Consistency Theorem and corollaries \\
 & (\texttt{sheaf\_consistency}, \texttt{compositionality\_sheaf}) \\
\texttt{ThreeViewExample.lean} & Concrete three-view site, design presheaf, \\
 & and main theorem (\texttt{three\_view\_main}) \\
\texttt{DesignCategory.lean} & Supporting categorical structures: \texttt{SystemDiagram}, \\
 & \texttt{CompatiblePair}, refinement adjunctions \\
\bottomrule
\end{tabular}
\end{center}

\end{document}